\documentclass[conference]{IEEEtran}
\IEEEoverridecommandlockouts
\usepackage{cite}
\usepackage{amsmath,amssymb,amsfonts}
\usepackage{algorithmic}
\usepackage{graphicx}
\usepackage{textcomp}
\usepackage{xcolor}
\usepackage{float}
\usepackage{hyperref}
\usepackage{amsmath}
\usepackage{amssymb}
\usepackage{array}
\usepackage{tablefootnote}

\def\BibTeX{{\rm B\kern-.05em{\sc i\kern-.025em b}\kern-.08em
    T\kern-.1667em\lower.7ex\hbox{E}\kern-.125emX}}
\begin{document}
\bstctlcite{IEEEexample:BSTcontrol}

\setlength{\columnsep}{0.24in} 

\title{Federated Learning and LLM-Driven Threat Intelligence for Zero Trust IoT Architecture}

\author{\IEEEauthorblockN{Amal Alshehri, Cihan Tunc}
\IEEEauthorblockA{\textit{Department of Computer Science \& Engineering} \\
\textit{University of North Texas}\\
\{Amal.Alshehri, Cihan.Tunc@UNT.edu\}}
}


\maketitle

\begin{abstract}
While the Internet of Things (IoT) has become essential, they introduced serious security and privacy challenges, especially for mission-critical environments. Legacy devices are vulnerable to viruses, data breaches, and unauthorized access, and updating these devices would be infeasibly costly. As a solution, this paper presents a Federated Learning and LLM-Driven Threat Intelligence for Zero Trust IoT Architecture, with FL for anomaly detection integrating privacy-preserving distributed learning, continuous identity verification, and LLM-driven autonomous threat response into a unified pipeline. Unlike existing solutions, our framework enforces Zero Trust at every communication layer via mutual TLS (mTLS) over MQTT, ensuring no device or message is implicitly trusted. Our experiments with Raspberry Pis and various sensors achieve an F1 score of 0.9091 and an ROC-AUC of 1.0, highlighting the effectiveness of the proposed framework in enabling privacy-preserving anomaly detection for resource-constrained IoT devices.

\end{abstract}


\section{Introduction}
The number of linked IoT devices are projected to reach to 39 billion  by 2030 with a compound annual growth rate of 13.2\% from 2025~\cite{sinha2025iotdevices}. 
According to a recent Fortune Business Insights analysis, the global IoT industry is expected to grow at a compound annual growth rate of 26.4\% to reach \$2.46 trillion in 2029~\cite{fortunebusinessinsights2026iot}.
Although IoT applications provide more convenient, automated, and intelligent lives, they have also introduced vulnerabilities. 
IoT devices are increasingly used in critical settings where data integrity and device reliability are crucial, such as industrial monitoring, smart buildings, and healthcare. Hence, IoT malware attacks experienced a 124\% increase globally in 2025 compared to 2024, highlighting a rapid growth in the number of IoT devices compromised and incorporated into botnets and other malicious campaigns~\cite{dexpose2026iotstats}.

As a countermeasure for the increasing attacks, the US government issued an executive order in May 2021 mandating that U.S. Federal Agencies adhere to National Institute of Standards and Technology (NIST) 800-207 as a necessary step for Zero Trust (ZT) implementation in response to the rising number of high-profile cyber attacks~\cite{rose2020zerotrust}.
The ZT Architecture (ZTA) states that all users, both inside and outside the company's network, must be verified, granted authorization, and routinely evaluated for security configuration and posture before gaining or maintaining access to apps and data~\cite{rose2020zerotrust}.

Federated Learning (FL) has emerged as a promising paradigm to address privacy and scalability challenges in distributed machine learning (ML) for each device to train a local model instead of sending raw sensor data to a central server. Only model changes, like gradients or weight deltas, are shared with a central aggregator~\cite{mcmahan2017communication}.  
In this study, we leverage FL for anomaly detection in IoT environments with limited resources, enabling 25 heterogeneous clients to collaboratively train a shared anomaly detection model without exposing raw traffic data, thereby preserving data privacy by design while maintaining high detection performance. 
After any abnormal activity, the detected event is forwarded to a large language model (LLM) that serves as the reasoning and explainability layer. The LLM analyzes the anomaly context together with relevant network and sensor metadata to generate an interpretable incident report and corresponding mitigation action. Depending on the severity level and risk assessment of the detected threat, the generated response can then be automatically translated into broker-level firewall policies and enforced directly on the MQTT infrastructure, enabling real-time threat isolation through blocking or restricting communication from malicious devices. 

\section{Related Work}

\subsection{Zero Trust Architecture (ZTA) in IoT environments} 

Li et al. proposed a blockchain-enabled zero-trust security framework (BasIoT) that leverages RSA digital signature-based authentication in 5G-IoT environments for the ``never trust, always verify'' principle. Each device must register on a private permission blockchain and authenticate every resource access request through a multi-step cryptographic verification~\cite{li2024future}.
Samniego and Deters have created Amatista, a blockchain-based middleware, to provide a reliable IoT network for an innovative hierarchical mining strategy that distributes the validating authority responsibilities both horizontally and vertically into various hierarchies of trust, going beyond the conventional blockchain mining process~\cite{8473444}. 
Zanasi et al. propose a flexible ZTA for heterogeneous IIoT environments, combining network micro-segmentation with a software-defined network (SDN) to enforce granular security policies while eliminating single points of failure through a peer-to-peer communication model with mutual authentication via WireGuard. Policy enforcement is delegated to individual network resources rather than a central controller, enabling fully decentralized operations with minimal overhead even on resource-constrained devices~\cite{zanasi2024flexible}. 

\subsection{Federated learning (FL) in IoT environments} 
Compared with centralized learning approaches, FL inherently enhances privacy and security by keeping locally generated data on end devices rather than transferring it to a central server~\cite{9709603}. 
Mothukuri et al. proposed an FL-based anomaly detection framework for IoT security using Gated Recurrent Units (GRUs) trained across decentralized virtual IoT instances without transferring raw data to a central server. Each client independently trains a local model using its private data and shares only the learned model weights with a central aggregator, which combines the updates to produce a global model that is subsequently shared with all clients. To further improve detection accuracy, an ensembler based on a random forest classifier combines the probability predictions from seven global ML models trained across different window sizes, achieving an average cross-validation accuracy of 99.5\%~\cite{mothukuri2021federated}.

Rey et al. propose a privacy-preserving FL for malware detection in IoT devices, evaluating both supervised multi-layer perceptron (MLP) classifiers and unsupervised autoencoder~\cite{rey2022federated}. 
They evaluated two FL aggregation algorithms: Mini-batch aggregation where the model is sent to the server after every mini-batch, and Multi-epoch aggregation where the model is trained for all epochs locally before transmission and reducing communication costs by $\sim$1300 times. 

\subsection{Large language models (LLM) in IoT environments}

Sarabi et al. explore the use of LLMs for analyzing textual data collected from Internet-wide network scans. The authors train a RoBERTa-based masked language model on large-scale banner data and fine-tune it using contrastive learning to generate stable embeddings representing network services and devices. These embeddings are then clustered using HDBSCAN to derive device fingerprints from HTTP banners. The results demonstrate that the approach can identify new IoT devices and server products not present in existing fingerprint databases such as Recog, highlighting the potential of LLMs for automated analysis of large-scale network measurement data~\cite{sarabi2023llm}. 
System that integrates lightweight LLMs fine-tuned on IoT-specific datasets to enable real-time anomaly detection and context-aware automated mitigation tailored for resource-constrained devices. 
A modular Docker-based architecture supports scalable and reproducible evaluation across different network conditions. Their experimental results demonstrate improved detection accuracy, reduced response latency, and enhanced resource efficiency compared to traditional security approaches, highlighting the effectiveness of LLM-driven autonomous security solutions for IoT environments~\cite{otoum2025llm}.
\section{Methodology}

\subsection{System Overview}
\begin{figure}[htbp]
    \vspace{-4mm}
    \centering
    \includegraphics[width=.85\columnwidth]{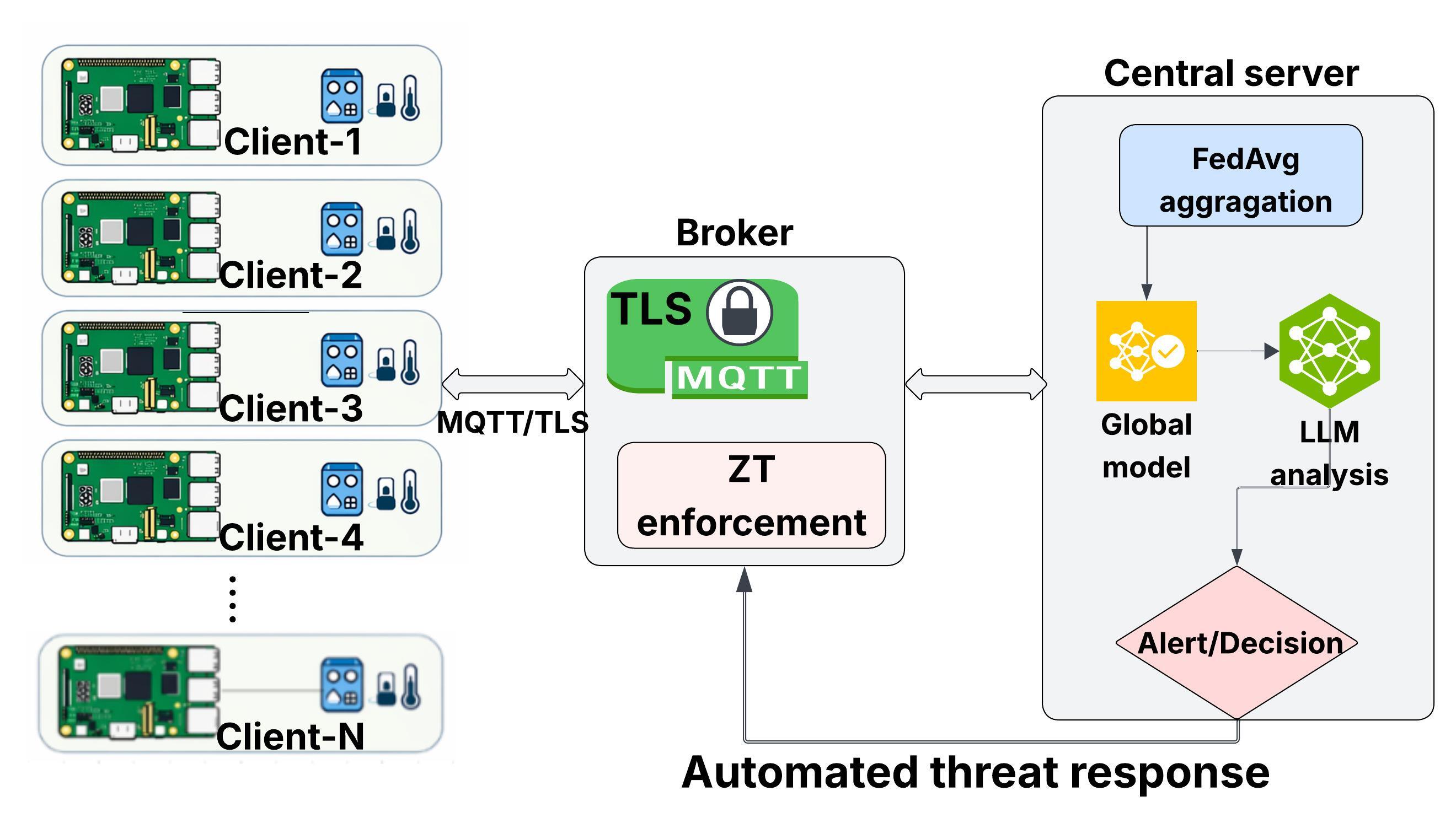}
    \caption{Architecture of the proposed FL framework for ZT in IoT.}
    \vspace{-4mm}
    \label{fig:system_overview}
\end{figure}

Our proposed FL-based and LLM-driven threat intelligence for ZTA for IoT ecosystem is presented in Fig.~\ref{fig:system_overview}. 
We assume a central MQTT broker is used for a distributed IoT environment with heterogeneous sensors and limited amount of processing capability. All devices communicate using MQTT over TLS 1.2 on port 8883 -- MQTT is a lightweight, broker-based publish–subscribe protocol operating over TCP/IP for resource-constrained devices~\cite{7968629}. When integrated with TLS, MQTT allows devices to publish sensor data securely. In this architecture, the MQTT broker functions as the central communication hub through which all device traffic is routed, making it an ideal point for monitoring and analysis. 
In order to broadcast the sensor data, we cannot rely on traditional perimeter-based security; because once a device enters the network, it is implicitly trusted, and one compromised device could compromise the system as a whole. Therefore, our system continually monitors network traffic, uses an FL autoencoder to detect suspicious activity, and uses a fine-tuned LLM to provide contextual threat intelligence, automatically and without human interaction implements network-level blocking. 
So, our architecture inherits ZT approach with no inherent trust in any device, user, or network flow, regardless of location or authentication history. 

\subsection{Feature Engineering}
The raw network traffic consists of over 80 features including constant data across MQTT/TLS communications or device-specific information that can lead to identity leakage. We applied a structured preprocessing pipeline to reduce the feature space to 22 discriminative features suitable for autoencoder training. During preprocessing, features are systematically removed based on four main criteria:
    Identity-related attributes (e.g., IP addresses, MAC addresses, and timestamps);
    Zero-variance features that remain constant across all samples;
    Fields that are consistently null in encrypted MQTT/TLS traffic; and
    Highly correlated features that provide redundant information.
The final feature set consists of six continuous variables that characterize network behavior, such as packet size, temporal features, and latency, along with sixteen binary features representing TCP flags, ports, TLS metadata, and packet type. The continuous features are then normalized to the [0, 1] range using a Min-Max normalization.

\subsection{FL-based Anomaly Detection}
In our proposed system, each IoT node trains a local model using its own data and shares only model parameters with the central server. These parameters are aggregated using the Federated Averaging (FedAvg) algorithm to produce a global model that captures the overall normal behavior across devices without exposing raw data.

\begin{figure}[htbp]
\vspace{-2mm}
    \centering
    \includegraphics[width=.9\columnwidth]{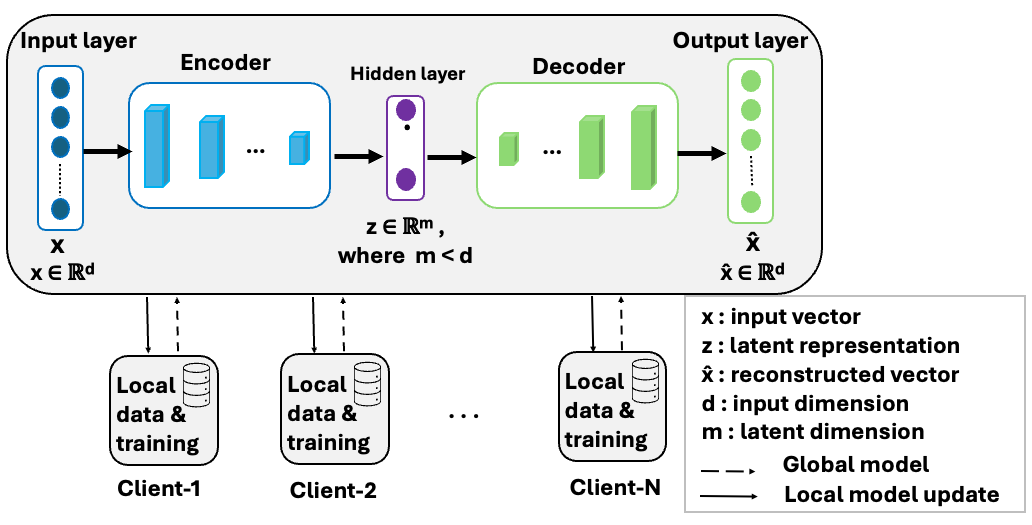}
    \caption{FL-based Autoencoder architecture.}
    \label{fig:Autoencoder_architecture}
    \vspace{-2mm}
\end{figure}

We focus on an Autoencoder instead of a supervised classifier to train the model only on normal traffic for expected network behavior.  
The proposed Autoencoder, Fig.~\ref{fig:Autoencoder_architecture}, consists of two components: an encoder and a decoder. The encoder compresses the input feature vector of dimension d=36 through three fully connected layers with dimensions 64, 32, and 16, progressively reducing the representation to a compact latent space of dimension m=16, with ReLU activation after every layer. The decoder reconstructs the original input from the latent representation of dimension m=16 through three fully connected layers, expanding progressively to dimensions 32, 64, and 36, with ReLU activation  after all layers except the  output layer. This bottleneck architecture forces the model to learn only the most essential features.

After training, the global model is evaluated on normal datasets to compute the anomaly detection threshold $\tau$. We set $\tau$ as the 95$^{th}$ percentile of the reconstruction MSE distribution over normal data, providing a 5\% theoretical false positive rate. Any packet exceeding $\tau$ is classified as ANOMALY and forwarded to the LLM for analysis. 

\subsection{LLM-Based Threat Intelligence Generation}
The Autoencoder produces a binary classification (Normal/Abnormal) along with a reconstruction error score; but, this information alone is insufficient for automated response. Effective mitigation requires identifying the nature of the attack and determining appropriate countermeasures. To bridge this gap, we integrate a fine-tuned LLM that converts anomaly detection outputs into structured threat intelligence reports, including actionable remediation commands. 
When an anomaly is detected, a structured report is generated with device information, reconstruction error and threshold comparison, and feature values. 
This report is passed to the LLM, which produces a structured threat intelligence and decision-making layer for ZTA enforcement. We used LoRA fine-tuning for efficient adaptation of a pre-trained model on a small domain-specific dataset without requiring full model retraining. The model is prompted with anomaly features and trained to produce a structured threat report containing verdict, severity, evidence, and automated response actions, closing the gap between anomaly detection and enforcement.

\subsection{Automated Enforcement}
The final stage of the system translates LLM outputs into automated response actions. Based on the severity and attack classification, appropriate mitigation measures are applied at the network level, such as blocking malicious IP addresses. 
This enables a fully automated detection-to-response pipeline, eliminating the need for manual intervention.

\section{Experimental Study} 

\subsection{Experimental Testbed}
To demonstrate our proposed framework, we created a realistic IoT testbed consisting of five Raspberry Pi devices with different sensors, an MQTT broker, and a centralized FL server, communicating over TLS (port 8883) using MQTT for a secure and authenticated data transmission. 
The broker serves as the central communication hub and hosts the network monitoring component, which captures all incoming and outgoing traffic using a custom packet sniffer. An IoT sniffer is implemented on the broker using Scapy. 
An FL server is deployed on a separate machine, responsible for coordinating training rounds and aggregating model updates from all devices. Each Raspberry Pi trains a local Autoencoder model on its own traffic data and sends only model weights to the server, preserving data privacy. 
To support large-scale evaluation, the testbed is extended with virtual devices generated using a statistical bootstrapping approach. This enables simulation of a larger and more diverse IoT environment without requiring additional physical hardware.
Our dataset, consisting of both real and synthetic network traffic, is as follows. 

\subsubsection{Normal Traffic}

Normal  data is collected in three sequential stages to capture IoT communication behavior across multiple Raspberry Pi devices.

\textbf{Baseline BME280 sensor data:}
The first stage consists of a homogeneous deployment with five Raspberry Pi devices (Pi1 - Pi5) with  BME280 environmental sensors operating under the same MQTT/TLS communication settings, publishing intervals, QoS levels, and payload structures. This stage establishes the baseline behavior. 

\textbf{Heterogeneous BME280 sensor data:}
We introduced controlled heterogeneity while still using BME280 sensors on all five Pis. Different communication configurations are assigned to each device to emulate realistic IoT deployment diversity; e.g., different MQTT publish intervals, payload formats, QoS levels, and MQTT topics. 

\textbf{AHT Sensor data:}
This stage collects traffic generated from AHT sensor deployments across the Raspberry Pi cluster to further diversify the environmental sensing patterns. 
During all stages, the broker-side network sniffer captures complete MQTT/TLS communication between devices and extracts protocol-level, timing, flow, TLS, MQTT, payload, and statistical traffic features directly from live network packets, and stores them in PostgreSQL for processing.

\subsubsection{Attack Traffic}

We first created a synthetic baseline representing normal MQTT/TLS IoT traffic and subsequently modified the selected network and protocol features to emulate malicious behavior. 
The generated attack scenarios are summarized as follows:
    \textbf{Port Scan: }
    Simulates TCP SYN half-open probing behavior using unexpected port activity, rapid packet transmission, high burst rates, reduced payload entropy, and disabled TLS encryption.
    \textbf{DDOS SYN Flood: }
    Represents volumetric SYN flooding behavior characterized by continuous SYN transmission, extremely small inter-arrival times, high burst activity, and strongly unidirectional traffic patterns.
    \textbf{TLS Downgrade: }
    Simulates degradation of secure MQTT/TLS communication by downgrading TLS~1.2 connections to weaker or absent encryption modes. 
    \textbf{Plain MQTT: }
    Represents unencrypted MQTT communication over port 1883 by disabling TLS-related features and modifying payload and timing to emulate plaintext MQTT traffic.

\subsubsection{Virtual Device Generation}
To generate realistic virtual IoT devices, we employ a coefficient of variation (CV)-based bootstrap framework that preserves both inter-device variability and intra-device stochastic behavior. Let $x_{j,f}$ denote the value of feature $f$ for sample $j$, where features are divided into continuous and binary sets. For each group $G$ (e.g., AHT, heterogeneous, BME280) and continuous feature $f$, the coefficient of variation is computed from dataset-level means as shown in Eq.~\ref{eq:CV}.

\begin{equation}\label{eq:CV}
\mathrm{CV}_{f,G} = \frac{\sigma_{f,G}}{\mu_{f,G} + \epsilon}
\end{equation}
where $\mu_{f,G}$ and $\sigma_{f,G}$ are the mean and standard deviation of per-device feature, and $\epsilon$ is a small constant for numerical stability. This value defines the allowable deviation of synthetic devices within the natural variability of real devices in the same hardware group. 
For each virtual device $v$, a feature-wise multiplicative shift is sampled once as in Eq.~\ref{eq:shift}:
\begin{equation}\label{eq:shift}
\delta_f^{(v)} \sim \mathcal{U}(-\mathrm{CV}_{f,G}, \mathrm{CV}_{f,G})
\end{equation}
which is applied consistently across all samples of that device to simulate its unique hardware characteristics. To preserve the joint distribution of features, samples are first drawn using bootstrap sampling as in Eq.~\ref{eq:boot}. 

\begin{equation}\label{eq:boot}
x_{j,f}^{\text{boot}} \sim \mathcal{D}_G
\end{equation}
where $\mathcal{D}_G$ is the pooled dataset of group $G$. 
For packet-level variability, independent Gaussian noise is added as in Eq.~\ref{eq:noise}. 
\begin{equation}\label{eq:noise}
\begin{array}{c}
     \eta_{j,f}^{(v)} \sim \mathcal{N}(0, \sigma_{\text{noise},f}^2), \\
\text{where }
\sigma_{\text{noise},f} = \max(0.02 \cdot \sigma_{f,G}, 10^{-6})
\end{array}
\end{equation}
ensuring that the magnitude of noise scales proportionally with the natural variability of each feature. 
The final generated feature value is computed as in Eq.~\ref{eq:final}:
\begin{equation}\label{eq:final}
\tilde{x}_{j,f}^{(v)} = \mathrm{clip}\left( x_{j,f}^{\text{boot}} \cdot (1 + \delta_f^{(v)}) + \eta_{j,f}^{(v)}, 0, 1 \right)
\end{equation}
where the clipping operation enforces the normalized range $[0,1]$. 
Binary and one-hot encoded features are excluded from this transformation and are directly copied from the bootstrapped samples as in Eq.~\ref{eq:binary}. 
\begin{equation}\label{eq:binary}
\tilde{x}_{j,f}^{(v)} = x_{j,f}^{\text{boot}}, \quad \forall f \in \mathcal{B}
\end{equation}
to preserve discrete constraints and ensure semantic validity. This formulation guarantees that synthetic devices remain statistically consistent with real devices, while introducing controlled variability bounded by $\mathrm{CV}_{f,G}$.

\subsubsection{LLM-Based Threat Analysis and Automated Response}
Llama 3.2-3B-Instruct was selected as the base model and fine-tuned using Low-Rank Adaptation (LoRA) on Google Colab via the Unsloth framework, which enables memory-efficient training through 4-bit quantization. The fine-tuning dataset consisted of 500 instruction–response pairs, balanced equally across the four attack classes: TLS downgrade, DDoS SYN flood, port scan, and plain MQTT (125 examples each). The ground-truth output is a JSON object consisting of the verdict, attack type, severity level (HIGH or CRITICAL), threat summary, risk assessment, a structured list of remediation actions, each with a shell command, and an automated response label (\texttt{REJECT\_CONNECTION} or \texttt{BLOCK\_SOURCE}).

The model was trained using the supervised fine-tuning trainer (\texttt{SFTTrainer}) for 3 epochs with a per-device batch size of 2 and gradient accumulation over 4 steps, resulting in an effective batch size of 8. The AdamW 8-bit optimizer was used with a learning rate of $2 \times 10^{-4}$, linear learning-rate decay, weight decay of 0.01, and 5 warm-up steps. Mixed-precision training was applied automatically, using BF16 when supported by the GPU hardware and FP16 otherwise.

For inference, the fine-tuned model is served through a Flask REST API hosted on Google Colab using an ngrok tunnel. The FL server submits anomaly reports to this API over HTTPS for threat analysis. When the autoencoder flags an anomaly, the feature vector and MSE ratio are formatted into the prompt template and submitted to the API, which returns a structured JSON decision. Upon receiving the decision, the automated response is enforced immediately. Anomalies classified as \texttt{BLOCK\_SOURCE} trigger \texttt{iptables} rules to drop all traffic from the malicious device IP at the broker.

The training hyperparameters were selected following established best practices for LoRA fine-tuning. A learning rate of $2 \times 10^{-4}$ was chosen to sufficiently adapt the LoRA weights to the structured JSON output format without degrading the base model's general language capabilities. The LoRA rank was set to $r=32$ with a scaling factor of $\alpha=64$, following the common convention of $\alpha = 2r$, which provides sufficient adapter capacity to learn the four attack classes while remaining parameter-efficient.

The per-device batch size was limited to 2 due to GPU memory constraints under 4-bit quantization. Gradient accumulation over 4 steps was therefore applied, resulting in an effective batch size of 8 to improve gradient stability. Training for 3 epochs was sufficient for convergence on the 500-sample dataset without noticeable overfitting.

\subsection{Experimental Results}

The global federated autoencoder was evaluated using a balanced dataset containing $957{,}620$ normal samples and $957{,}620$ attack samples, with $239{,}405$ samples generated for each attack category, resulting in a $1{:}1$ class distribution. The anomaly detection threshold $\tau$ was defined as the $95^{\text{th}}$ percentile of the reconstruction error distribution computed from all normal traffic samples. 
The anomaly detection performance is summarized in Table~\ref{tab:ae_results}. The proposed FL-based autoencoder achieved identical performance across all evaluated attack classes. 
Each attack category achieved an accuracy of 0.96, precision of 0.83, recall of 1.00, F1-score of 0, and ROC-AUC of 1.00. The perfect recall confirms that all attack samples produced reconstruction errors above the threshold to be correctly detected. The precision of 0.8333 reflects the fixed 5\% false positive rate introduced by the 95th percentile threshold, where a small proportion of normal samples are incorrectly flagged by design. The ROC-AUC of 1.0000 confirms perfect separation between normal and attack reconstruction error distributions across all attacks.  

\begin{table}[htbp]
\vspace{-2mm}
\centering
\caption{Autoencoder detection performance.}
\label{tab:ae_results}
\scriptsize
\begin{tabular}{lcccccc}
\hline
Class & Samples & Acc & Prec & Rec & F1 & Spec \\
\hline
Normal      & 957620 & -- & -- & -- & -- & 0.950 \\
DDoS SYN    & 239405 & 0.960 & 0.833 & 1.000 & 0.909 & -- \\
Plain MQTT  & 239405 & 0.960 & 0.833 & 1.000 & 0.909 & -- \\
Port Scan   & 239405 & 0.960 & 0.833 & 1.000 & 0.909 & -- \\
TLS Down.   & 239405 & 0.960 & 0.833 & 1.000 & 0.909 & -- \\
\hline
Macro Avg   & 957620 & 0.960 & 0.833 & 1.000 & 0.909 & -- \\
Overall     & 1915240 & 0.975 & 0.952 & 1.000 & 0.976 & 0.950 \\
\hline
\end{tabular}
\vspace{-1mm}
\end{table}
To further evaluate robustness under realistic IoT class imbalance, additional experiments were performed using highly imbalanced datasets while keeping the normal dataset fixed at $957{,}620$ samples and progressively reducing the number of attack samples. The evaluation considered imbalance ratios ranging from $1{:}1$ to $100{:}1$. Across all attack types and imbalance settings, the proposed model consistently achieved perfect recall ($1.0000$), indicating that no attacks were missed even under severe imbalance conditions. Although precision and F1-score decreased as the imbalance ratio increased due to the fixed false positive rate introduced by the p95 thresholding strategy, ROC-AUC remained $1.0000$ across all experiments, demonstrating perfect reconstruction-error separability between normal and malicious traffic. These results confirm the robustness of the proposed FL-based anomaly detection under highly imbalanced IoT security scenarios.

\begin{figure}[htbp]
\centering
\includegraphics[width=0.95\linewidth]{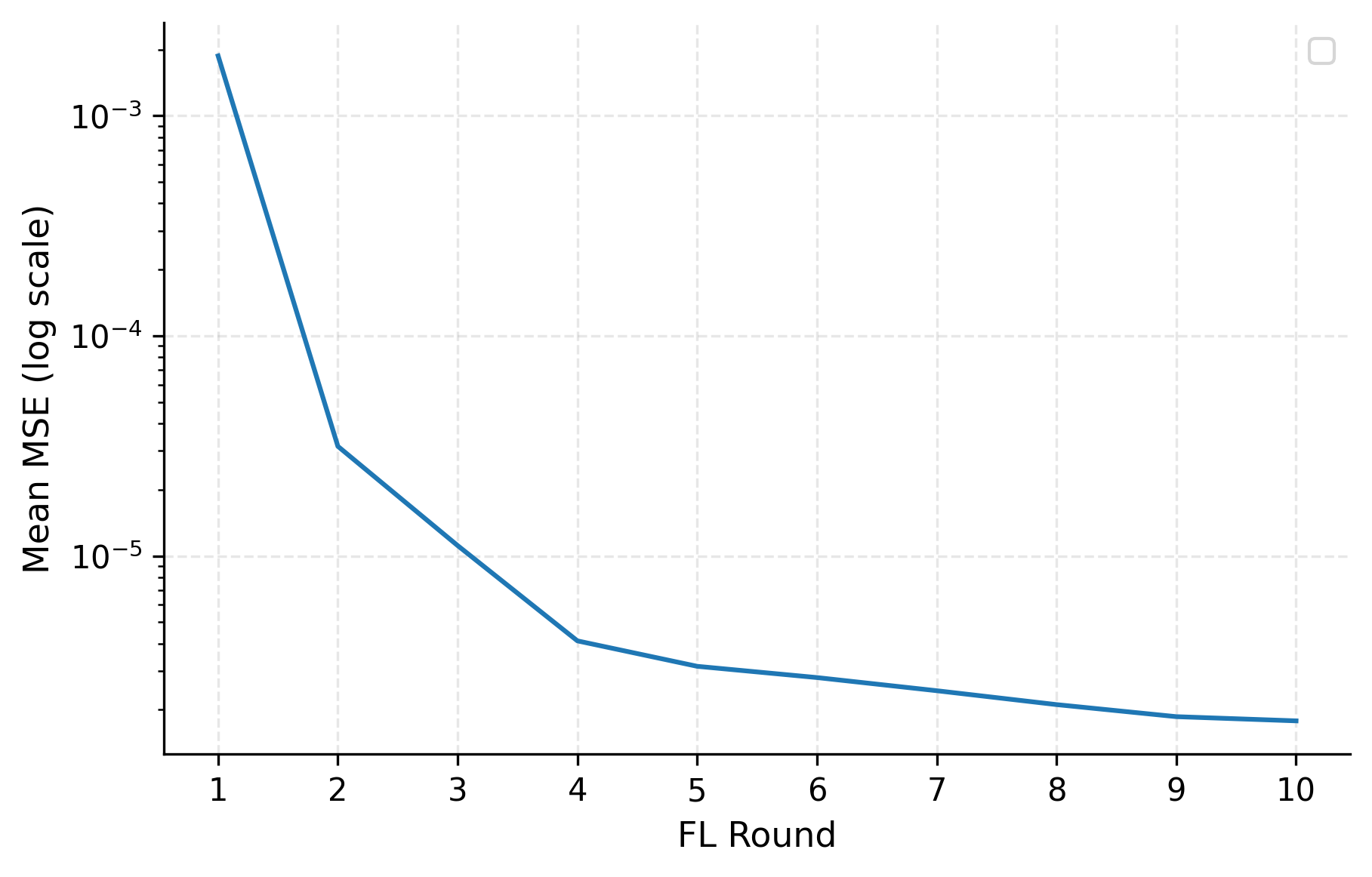}
\caption{FL convergence of the global autoencoder across 10 comm. rounds.}
\label{fig:fl_convergence}
\vspace{-4mm}
\end{figure}

The convergence behavior of the federated autoencoder during collaborative training is given in Fig. ~\ref{fig:fl_convergence}. The global reconstruction loss decreases rapidly during the initial communication rounds and gradually stabilizes, indicating successful convergence of the FedAvg training process.  

The computational overhead associated with FL is summarized in Table~\ref{tab:fl_resource} showing the proposed FL maintains relatively low server-side overhead while supporting distributed training across $25$ heterogeneous clients. The average FL round duration was $597.12$ seconds, while the mean FedAvg aggregation time was $2530.81$\,ms. The average communication overhead per round was $2408.86$\,KB, demonstrating that collaborative learning can be achieved with moderate communication cost.

\begin{table}[htbp]
\vspace{-2mm}
\centering
\caption{FL resource req. averaged across all comm. rounds.}
\label{tab:fl_resource}

\scriptsize

\begin{tabular}{p{3.5cm} p{1.6cm} p{2.2cm}}
\hline
\textbf{Metric} & \textbf{Value} & \textbf{Description} \\
\hline
Mean round duration  & 597.12 s & Average FL round execution time \\
Mean FedAvg time  & 2530.81 ms & Average server aggregation time \\
Mean communication/round  & 2408.86 KB & Mean server and client communication traffic size \\
Mean server CPU  & 0.48\% & Average server CPU utilization \\
Peak server RAM  & 383.61 MB & Maximum observed server memory usage \\
\hline
\end{tabular}
\vspace{-2mm}
\end{table}

Client-side resource requirements measured on the Raspberry Pi devices are presented in Table~\ref{tab:ae_resource} with an average inference latency of only $0.0065$\,ms per sample, demonstrating suitability for lightweight real-time IoT deployment. The average Pi CPU utilization during local training was $46.38\%$, while the maximum memory usage reached $553.12$\,MB.

\begin{table}[htbp]
\vspace{-3mm}
\centering
\caption{Autoencoder client-side resource requirements averaged across all FL clients and rounds.}
\label{tab:ae_resource}

\scriptsize

\begin{tabular}{p{2.3cm} p{1.2cm} p{3.2cm}}
\hline
\textbf{Metric} & \textbf{Value} & \textbf{Description} \\
\hline
Mean AE inference  & 0.0065 ms & Per-sample inference latency \\
Mean train time/round  & 115.26 s & Local client training time \\
Mean train time/epoch  & 23.05 s & Local epoch duration \\
Mean client CPU  & 46.38 \% & Average Raspberry Pi CPU usage \\
Peak client RAM  & 553.12 MB & Maximum observed client memory usage \\
\hline
\end{tabular}

\end{table}

The classification performance is summarized in Table~\ref{tab:llm_results}. The LLM achieved perfect precision and recall ($1.0000$) across all four attack classes, correctly identifying the attack category and severity level for every anomalous sample. 
The overall F1-score of $0.9231$ was slightly reduced due to false positives due to the autoencoder thresholding. These occur when benign samples exceed the p95 reconstruction error threshold and are subsequently classified as malicious by the LLM. 

\begin{table}[htbp]
\centering
\caption{LLM classification performance across attack categories.}
\label{tab:llm_results}

\scriptsize

\begin{tabular}{lcccc}
\hline
\textbf{Attack Type} & \textbf{Acc} & \textbf{Prec} & \textbf{Rec} & \textbf{F1} \\
\hline
Port Scan      & 1.0000 & 1.0000 & 1.0000 & 1.0000 \\
DDOS Attack    & 1.0000 & 1.0000 & 1.0000 & 1.0000 \\
TLS Downgrade  & 1.0000 & 1.0000 & 1.0000 & 1.0000 \\
Plain MQTT     & 1.0000 & 1.0000 & 1.0000 & 1.0000 \\
\hline
\textbf{Macro-average} & 1.0000 & 1.0000 & 1.0000 & 1.0000 \\
\textbf{Overall}       & 0.9600 & 0.8571 & 1.0000 & 0.9231 \\
\hline
\end{tabular}
\vspace{-4mm}
\end{table}

The measured LLM reasoning latency is as follows: The average end-to-end latency was $41.41$ seconds, including HTTP communication, Colab T4 inference, ngrok network overhead, and JSON response generation. The latency remained relatively stable across all calls, with a standard deviation of $1.42$ seconds. And, we observed a minimum latency of $39.32$ and maximum latency of $45.39$ seconds. 

The latency across different attacks is summarized in Table~\ref{tab:llm_attack_latency}. The mean response ranged from $40.24$ seconds for plain MQTT attacks to $43.26$ seconds for DDOS attacks. The relatively small latency variation indicates stable LLM inference behavior across diverse attack scenarios.

\begin{table}[htbp]
\vspace{-2mm}
\centering
\caption{LLM latency by attack type.}
\label{tab:llm_attack_latency}

\scriptsize

\begin{tabular}{p{2.8cm} c c c}
\hline
\textbf{Attack Type} & \textbf{Mean (s)} & \textbf{Median (s)} & \textbf{p95 (s)} \\
\hline
Port Scan & 40.38 & 40.34 & 41.56 \\
DDOS Attack & 43.26 & 43.45 & 45.28 \\
TLS Downgrade & 41.76 & 41.70 & 42.24 \\
Plain MQTT & 40.24 & 40.29 & 40.55 \\
\hline
\end{tabular}
\vspace{-2mm}
\end{table}

Overall, the results demonstrate that combining FL-based anomaly detection with domain-specific LLM reasoning enables accurate and explainable threat classification. 

\begin{figure}[h]
\centering
\includegraphics[width=0.9\linewidth]{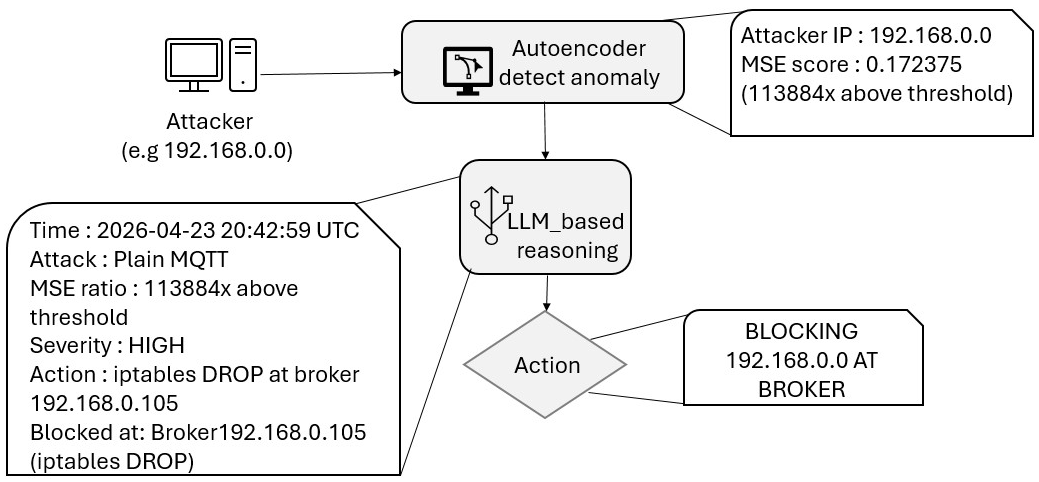}
\vspace{0.3cm}
\caption{End-to-end ZT-FL enforcement pipeline. }
\label{fig:enforcement_pipeline}
\end{figure}

The complete enforcement workflow is illustrated in Fig. ~\ref{fig:enforcement_pipeline}, including anomaly detection, LLM reasoning, MQTT alert generation, and automated firewall generation. 
The proposed framework successfully blocked all attacker IPs within a single evaluation cycle. The generated \texttt{iptables} firewall rules were immediately applied on the broker machine. Additionally, all enforcement actions were recorded in the audit log with associated metadata, including device IP address, attack type, reconstruction error ratio, and timestamp. The complete end-to-end detection and response pipeline required approximately $44$ seconds due to remote LLM inference. But, once a malicious decision was produced, the enforcement operation itself completed in less than one second.

\section{Conclusion}

This work demonstrates that combining FL with LLM-driven reasoning enables a practical, scalable, and fully autonomous Zero Trust defense for IoT environments, capable of detecting, interpreting, and mitigating threats. The proposed framework demonstrates that Zero Trust security can be effectively implemented in IoT environments by integrating FL, LLM-based threat intelligence, and automated enforcement into a unified pipeline. The federated autoencoder achieved perfect recall across all evaluated attacks while preserving data privacy by training exclusively on local device data. The fine-tuned LLM successfully translates anomalies into accurate, actionable security decisions with automated response without human intervention.

\bibliographystyle{IEEEtran}
\bibliography{references}

\end{document}